\documentstyle[12pt,amssym,aasms4,tighten]{article}

\lefthead{Park et al.}
\righthead{Real-Time Optical Flux Limits From Gamma-Ray Bursts Measured By \\
	The GROCSE Experiment}

\begin{document}

\title{Real-Time Optical Flux Limits From Gamma-Ray Bursts Measured By \\
	The GROCSE Experiment}

\author{H. S. Park\altaffilmark{1}, 
	E. Ables\altaffilmark{1},
	D. L. Band\altaffilmark{5},
	S. D. Barthelmy\altaffilmark{3,7},
	R. M. Bionta\altaffilmark{1},
	P. S. Butterworth\altaffilmark{3},
	T. L. Cline\altaffilmark{3},
	D. H. Ferguson\altaffilmark{6},
	G. J. Fishman\altaffilmark{4},
	N. Gehrels\altaffilmark{3},
	K. Hurley\altaffilmark{8},
	C. Kouveliotou\altaffilmark{4},
	B. C. Lee\altaffilmark{2},
	C. A. Meegan\altaffilmark{4},
	L. L. Ott\altaffilmark{1} and
	E. L. Parker\altaffilmark{1}}


\altaffiltext{1}{Lawrence Livermore National Laboratory, Livermore, CA 94550} 
\altaffiltext{2}{Dept. of Physics, University of Michigan, Ann Arbor, MI 48109} 
\altaffiltext{3}{NASA/Goddard Space Flight Center, Greenbelt, MD 20771}
\altaffiltext{4}{NASA/Marshall Space Flight Center, Huntsville, AL 35812}
\altaffiltext{5}{CASS 0424, University of California, San Diego, La Jolla, CA 92093}
\altaffiltext{6}{Dept. of Physics, California State University at Hayward, Hayward, CA 94542}
\altaffiltext{7}{Universities Space Research Association}
\altaffiltext{8}{Space Sciences Laboratory, University of California, Berkeley, CA 94720-7450}


\begin{abstract}
The Gamma-Ray Optical Counterpart Search Experiment (GROCSE) presents new
experimental upper limits on the optical flux from gamma-ray bursts (GRBs).
Our experiment consisted of a fully-automated very wide-field opto-electronic
detection system that imaged locations of GRBs within a few seconds of
receiving trigger signals provided by BATSE's real-time burst coordinate
distribution network (BACODINE). The experiment acquired ~3800 observing
hours, recording 22 gamma-ray burst triggers within $\sim$30 s of the start 
of the
burst event. Some of these bursts were imaged while gamma-ray radiation was
being detected by BATSE. We identified no optical counterparts associated with
gamma-ray bursts amongst these events at the m$_V$ $\sim$ 7.0 to 8.5 sensitivity
level. We find the ratio of the upper limit to the V-band optical flux, 
F$_\nu$, to the gamma-ray fluence, $\Phi_\gamma$, from these data to be
$2 \times 10^{-18} < F_\nu/\Phi_\gamma < 2 \times 10^{-16}$.
\end{abstract}


\keywords{gamma rays: bursts}


%

\section{Introduction}

Cosmic GRBs are arguably the most intriguing phenomena in modern high energy
astrophysics. Resolving the mystery of the origin of gamma-ray bursts will
probably require the discovery of counterpart radiation in other wavebands,
particularly in the optical band.  Detailed study of such a counterpart might
reveal the physical mechanisms of the GRB process while even a single
unambiguous distance measurement could determine the burst distance scale. At
the very least, upper limits at other frequencies constrain sources'
multi-wavelength characteristics and thus possible emission mechanisms. Given
the possibility that GRBs may represent hitherto unexplored physical regimes,
it is not surprising that several of near simultaneous counterpart searches
are in progress (see Hudec 1995a, b and references therein). Here we report on
the results of two years of operation of the Gamma-Ray Optical Counterpart
Search Experiment (GROCSE).

Counterpart emission can be described as flaring, fading, or quiescent
(Schaefer 1994). Flaring emission would be simultaneous with the burst
observed in the gamma-ray band. One approach to observing the flaring emission
is to scan the sky independent of the gamma-ray detector. This is the strategy
of the Ondrejov photographic network (Hudec 1993; Greiner et al. 1992, 1994,
and 1995; Hudec et al. 1995b), and the Explosive Transient Camera
(ETC--Vanderspek et al. 1994). It was also the intended strategy of the
ill-fated High Energy Transient Explorer (HETE--Ricker et al. 1988).
Alternatively, a detector can respond rapidly to a burst trigger from a
gamma-ray instrument, catching a burst in its later phase. This approach was
used by GROCSE and has been adopted by other operational systems as well: the
BATSE-COMPTEL-NMSU network (McNamara 1996), the Livermore Optical Transient
Imaging System (LOTIS) (Park 1997), and the Department of Defense 
GEODSS system. Burst
coordinates are currently distributed in near real-time by the BATSE
CO-ordinate DIstribution NEtwork (BACODINE--Barthelmy et al. 1994), 
which monitors
BATSE data on board and responds to bursts by calculating and distributing
preliminary burst positions.

Optical emission may come from the burst source or its environment after the
gamma-ray emission ends. Such emission, if luminous enough, would allow
detection of a source not previously present in the seconds-to-days after a
GRB and which would subsequently fade. There are currently a large number of
optical and radio experiments, including those searching for flaring emission
mentioned above. These search experiments may be characterized by the energy
band searched, the depth of the search (e.g. the limiting magnitude), and the
time between the burst and the search (McNamara 1995). The depth-of-search
improves as the GRB-to-optical observation time interval increases due to lag
times necessary to compute minimum-field error boxes and to arrange
observations on large telescopes. For example, the typical limiting magnitude
on time scales of hours is m$_V \sim$ 8, while it improves to 
m$_V \sim$ 22 on time scales longer than 1 day. Most of these searches 
for fading emission use BACODINE positions. If a burst location also happens 
to fall within the COMPTEL or EGRET fields-of-view, a more precise position 
can be distributed in a matter of days. On similar time scales, the
Interplanetary Network, which currently consists of only one long baseline
(Earth-Ulysses--Hurley et al. 1994, 1997), provides arcs only a few arcseconds wide.

The smaller error boxes obtained at longer time scales have been searched by
a variety of powerful telescopes in different wavebands. Objects potentially
of interest will be found if a small region of the sky is searched
sufficiently deep, and therefore the issue is whether any likely sources are
found. Such searches might detect an interesting object within the error boxes
if bursts originate within the Galaxy. If bursts instead originate at
extragalactic distances, then a galaxy is expected in the error box. In
particular, a bright burst is presumably relatively nearby and its host galaxy
should be bright. Bright bursts can usually be localized to smaller error
boxes, diminishing the probability of an unrelated bright galaxy in the error
box.

There are no firm predictions of expected optical emission because the origin
and nature of GRBs are uncertain. Schaefer's report of optical transients on
archival photographic plates (Schaefer 1981; Schaefer et al. 1984) led to a
number of theoretical models that in general attempted to explain the apparent
magnitudes of these optical transients. These theories included reprocessing
gamma rays in a stellar companion's atmosphere (London\& Cominsky 1983; London
1984; Rappaport \& Joss 1985; Melia, Rappaport, \& Joss 1986; Cominsky, London,
\& Klein 1987) or in an accretion disk (Epstein 1985; Melia 1988). Other
theories attributed optical-ultraviolet emission to processes in a neutron
star magnetosphere (Liang 1985; Katz 1985; Bisnovatyi-Kogan \& Illarionov 1986;
Sturrock 1986; Ruderman 1987; Hartmann, Woosley, \& Arons 1988; Hameury \&
Lasota 1989; Ho \& Epstein 1989, Dermer 1990) and assumed source distances of
order 100 pc. The latter theories were largely invalidated by the BATSE
observations, which placed large lower distance limits and instead support
Galactic halo or cosmological models. Ford \& Band (1996) found by simple
extrapolation of burst spectra that, at any given site, flaring emission would
be observable from only a few bursts per year with m$_V$ $\sim$ 10 to 15. 
The brighter predicted optical fluxes result from bursts with soft gamma-ray 
spectra. If we instead assume the low energy photon spectrum is 
F$_\gamma \propto E^{1/3}$, as expected for synchrotron emission from an 
electron distribution with a low-energy cutoff
(Katz 1994; Tavani 1996a, b), then the predicted distribution of optical
fluxes will be shifted towards fainter magnitudes.

A number of theories suggest that fading optical emission may be observable
as a result of reprocessing of burst radiation by the medium surrounding the
burst source: Jennings (1983) considered H$\alpha$ line radiation from resonant
scattering and recombination, while Chevalier (1986), Schaefer (1987), and
Katz \& Jackson (1988) investigated dust scattering of radiation from an
optical transient. These models predict that ionizing radiation is reprocessed
to optical radiation in a manner thus far unobserved. The fading emission is
proportional to the burst fluence and the density of the reprocessing region.
For sufficiently large optical emission, the density must be 
n $\sim 10^5$ $cm^{-3}$ (Band \& Hartmann 1992). These models were 
constructed within the paradigm of local burst sources and thus do not 
provide usable optical predictions, if one
accepts the current paradigm of halo cosmological distances.

In cosmological fireball models, a relativistically expanding fireball
radiates at the shocks formed either as the ejecta within the fireball impact
the surrounding medium (Rees \& Meszaros 1992; Meszaros \& Rees 1993; Meszaros,
Rees, \& Papathanssiou 1994; Katz 1994; Sari, Narayan, \& Piran 1996) or as a
consequence of inhomogeneities within the expanding fireball (Rees \& Meszaros
1994; Paczynski \& Xu 1994; Papathanssiou \& Meszaros 1996). In these models,
the spectrum extends from the gamma ray to the optical band during the burst
itself (e.g. Papathanssiou \& Meszaros 1996). In addition, the optical
emission may last for hours after the end of the gamma ray emission. The
predicted optical flux and its temporal evolution are highly model-dependent.
For example, in considering fading optical emission for a variety of models,
Meszaros \& Rees (1997) find $9 \leq m_V  \leq  19$ which fades as $A \log t(s)$
where $3.75  \leq A  \leq 15$. These cosmological models will thus undoubtedly
accommodate upper limits and possible detection from GROCSE and successor 
experiments.

The BATSE gamma-ray detectors provide burst positions localized only within
wide angular limits, $1 \sim 10^\circ$, depending on burst brightness 
and duration (Meegan 1996). A system that seeks to detect an optical counterpart
near-simultaneously with the GRB must be on-location within a few seconds and
image over a field-of-view roughly 15$^\circ$ across. GROCSE was the first
operational instrument satisfying these criteria. GROCSE collected images of
burst error boxes shortly after the bulk of the gamma-ray emission, and
therefore the GROCSE results place the strongest constraints yet on optical
counterparts of fading GRB emission on time scales of tens of seconds from the
GRB peak.

\section{GROCSE Instrumentation}

The basic requirements to search for GRB optical counterparts are: receiving
and processing in near-real time GRB coordinates; a $> 15^\circ$ field of view
telescope system to cover the large angular uncertainty of the BATSE gamma-ray
detectors onboard the Compton Gamma-Ray Observatory; and a rapidly-slewing
telescope mount capable of slewing to the GRB field within a few seconds.
GROCSE received near-real time GRB coordinates from BACODINE which uses
real-time data from the satellite, computes burst coordinates using the
weighted triangulation method, and transmits the information via the Internet
(Barthelmy 1994). The "internal socket" protocol we employed for data
communication between NASA/Goddard Space Flight Center (GSFC) and our
observation site at the Lawrence
Livermore National Laboratory (LLNL) established special dedicated links to
transmit and receive pre-formatted data packets. We verified the connection
once per minute during observations by sending test packets. The link has
transmitted data reliably for over 3 years. The average delay from the BATSE
trigger to receipt by the GROCSE pointing and control software is $\sim$5.5 s.
Our electro-optical sensor and data recording equipment derived from a system
originally constructed for a low-Earth orbit satellite tracking program (Park
1990). It featured a very large, fast-slewing alt-azimuth telescope mount and
unique wide-field optics. The hardware is shown and illustrated
schematically in Figure 1. A fish-eye lens constructed of solid blocks of
concentric spherical elements produces uniform spot sizes across the entire
lens field of view of 60$^\circ$. The effective aperture of this lens is 
89 mm and the focal length is 250 mm. We covered the large spherical focal 
area ($>$ 500 cm$^2$) with 23 segments of custom fiber optic reducers 
(3.8:1 reduction ratio). The front of each fiber bundle was machined to match
the curvature of the spherical lens for best focus. Each focal reducer mapped 
an $8 \times$ 11$^\circ$  field of view onto an $8 mm \times 13 mm$ flat CCD 
area. 

The schematic of the data acquision system is shown in Figure 2. The 23 
cameras were clocked synchronously by a master clock and timing
distribution box. The gain of each intensifier was computer controlled through
a CAMAC interface. The exposure duration for all GROCSE images was set by the
length of the intensifier gate pulse to 0.5 s. The entire lens and camera
assembly was mounted on a Contraves, Goertz, Inc. computer-controlled inertial
guidance indexing table. The mount provided a maximum angular slew rate of
1000/sec with pointing precise to $\sim$ 1 arcsec. Our data collection system was 
hosted by a SUN 4/330 computer. The camera output was directed to an 8 bit 
digitizer by means of a 23 channel video multiplexer. Following digitization, 
these data were collected and formatted by a Datacube image processing system, 
time-tagged using a WWVB clock, and then stored on disk. In addition, all GRB 
events were archived to Exabyte tape. 
The Sun 4/330 host also provided pointing commands to the Contraves, Goertz,
Inc. mount via a GPIB interface. In addition, the computer performed various
housekeeping controls, such as monitoring the precipitation detector for
indications of rain or fog and closing a weathertight clam shell over the
instrument during periods of daylight or inclement weather.
The on-line software automatically activated the instrument shortly after
sunset and reestablished the connection to BACODINE. While awaiting GRB
triggers, GROCSE collected data across the entire sky every 30 minutes,
recording suitable sky background data for analysis of any GRB events across
the entire unobstructed sky. We limited our observations to more than 30$^\circ$
above the horizon because buildings surrounding the system, which was
necessarily on-site at LLNL, blocked the view at lower angles. This "sky
patrol" was interrupted whenever  BACODINE sent GRB coordinates. If such a
coordinate set was within the field of regard of the telescope system and not
within 30$^\circ$ of the Moon, GROCSE was slewed at rapid rate to the 
location of the GRB candidate and images were recorded for 20 minutes after 
the trigger.  GROCSE  then returned to "sky patrol" mode until either 
another GRB occurred or dawn broke. The software was programmed to 
deactivate the camera and close the clam shell before dawn. GROCSE 
operated between 1994 January and 1996 June for 3,800 hours (Lee 1997). The
instrument operated approximately 52\% of available night-hours. GROCSE
recorded 22 burst triggers during the period of operation.


\section{GROCSE Gamma-Ray Burst Observations}

The basic analysis strategy was to search the resulting images of GRB
location for "new" star-like objects that do not appear in the star catalogues
or in the background sky patrol images. The task of identifying stars in the
imagery was complicated by the noisy intensifier data, the necessity to reject
ion events, and differences in field distortion between all detectors due to
variations in the microchannel plate bundle uniformities. We utilized a
program that automatically matched known positions of bright stars to stellar
locations from GROCSE images. Ambiguity occasionally existed, such as in field
areas suffering from high microchannel plate distortion, in crowded fields, or
where an cosmic ray event may have occurred. In these cases the operator
examined carefully the star image in question to either assign a bright star,
or if the datum in question subtended only one pixel, reject the point as an
cosmic ray event. We found the method to be very reliable except for double 
stars with known separations between 50" and 300", in which case very careful
comparison of stellar coordinates with image pattern was required. All images
from 13 GRB triggers was examined in this fashion. None of the images
contained any evidence for "new" star like objects.

We derived upper limits to in-band fluxes of gamma-ray bursts shortly after
outburst peaks by computing the sensitivity of GROCSE from observations of
stars obtained in the GRB field. This process circumvented the need to
radiometrically calibrate the individual sensors each night. Our data yield a
completeness limit across observed gamma-ray burst events of 
m$_{V complete}$
$\sim$ $6.75\pm0.25$ 1-$\sigma$, which corresponds to a flux density 
$F_\nu (5500\AA)$ = $7.5\pm0.7 \times 10^{-23}$
ergs-cm$^{-2}$-s$^{-1}$-Hz$^{-1}$. 
Our limiting magnitude is m$_{V limit}$ $\sim$ $7.4\pm0.4$ 1-$\sigma$,
or a respective flux density $F_\nu (5500 \AA)$ = 
$4.1\pm0.8 \times 10^{-23}$ ergs-cm$^{-2}$-s$^{-1}$-Hz$^{-1}$. 
Note that the  completeness magnitude is defined as the
level for which all stars are observed while the limiting magnitude is the
brightness where only half the stars are observed.

The GROCSE GRB data are summarized in the first seven columns of Table 1.
Column 1 lists the GRB name from their recorded UTC date. The BATSE trigger
number is listed in column 2. We analyzed 13 of the 22 total GRBs. The other
events were rejected due to cloud conditions or a error in BACODINE position.
Our measured limiting visual magnitude m$_V$ is shown in column 3, 
then converted to frequency-dependent flux at 5500$\AA$ $F_\nu (5500\AA)$ 
as listed in column 4.
Column 5 shows the time delay td from the beginning of the burst to the
collection time of the optical images and column 6 specifies the gamma-ray
burst duration time (T$_{90}$) measured by BATSE. Columns 5 and 6 can be
compared to see if a given GRB was imaged optically while the gamma-ray 
event was still in progress. Column 7 gives the percent of the BATSE 
3-$\sigma$ error box imaged by GROCSE.

We have selected three BATSE triggers, GRB951117, 951124, \& 951220 from our
observations to show some of the event characteristics and analysis. The IPN
annuli are available for these events which give smaller positional error.
Figure 3-5 shows BATSE light curves for the three events and the BATSE 
3-$\sigma$
error circle for the three events in question, each superimposed on the GROCSE
rectangular field. A systematic error of 1.6$^\circ$ is included. 
Burst number, date,
limiting magnitude, and the fraction of the BATSE error circle superscribed by
the GROCSE frame are given for each event.

We considered the effect of the uncertainties in stellar effective temperatures,
variable stars, differences between cameras in our focal plane array, sky 
background, and star identification errors. Stars with a range of effective 
temperatures were observed across the approximately 0.32-0.70 $\mu$m in-band 
sensitivity region of the S-20 photocathodes. Stars of different spectral 
types, when combined with the wavelength-dependent instrument response 
function, will have varying ratios, $\Delta$m$_{V T}$, of apparent m$_V$ 
at 5550 $\AA$ to in-band flux. This quantity was computed for a variety of 
effective temperatures where
\begin{equation}
\Delta m_{V T} = -2.5 \times log 
\Biggl ( 
{
  {
    {   
      \int^{\lambda_r}_{\lambda_b} 
      B_\lambda (T) \enspace
      \tau_{inst} \enspace
      \tau_{atm} \enspace
      \eta \enspace
      d\lambda
    }
    \quad
    /
    \quad
    { 
      B_{5500 \AA} (T, \theta)
    }
  }
  \over
  {
    {   
      \int^{\lambda_r}_{\lambda_b} 
      B_\lambda (10,000K) \enspace
      \tau_{inst} \enspace
      \tau_{atm} \enspace
      \eta \enspace
      d\lambda
    }
    \quad
    /
    \quad
    { 
      B_{5500 \AA} (10,000K, \theta)
    }
  }
}
\Biggr ) \enspace .
\eqnum{1}
\end{equation}
Here, $\theta$ is the zenith angle, B$_\lambda$ (T) is the blackbody flux 
at wavelength $\lambda$ for
stellar effective temperature T, $\tau_{inst}$ is the wavelength-dependent instrument throughput to the sensor, $\tau_{atm}$ is the wavelength and zenith
angle-dependent atmospheric transmission (Allen 1976), and $\eta$ is the S-20 
photocathode quantum
efficiency. Given, as we shall see, that the magnitude spread in our
observations is considerably larger than the color effect component, we
approximated stellar energy distributions with Planck distribution for ease of
calculation. These differences amounted to no more than 
$\Delta$m$_{V T}$ = $\pm$0.13, 1-$\sigma$,
for stars with effective temperatures ranging from 3000 K to 20,000 K with a
distribution of equal numbers of stars per unit linear temperature. This is
clearly a very conservative upper limit to temperature sensitivity because the
vast majority of all stars observed to a limiting magnitude m$_V <$ 7.5 have
effective temperatures between 4,000 K and 15,000 K. The dependence of 
$\Delta$m$_{V T}$ on zenith angle is small compared with the effect of the 
rapidly changing S-20 photon response with wavelength.

There are of order 10$^4$ variable stars listed in Kholopov (1987) distributed
across $\sim$10$^4$ square degrees of sky, or of order one variable star 
per square degree. Including as it does all known variable stars, the 
Kholopov sample is
increasingly complete for stars of brighter apparent magnitude and wider
brightness variation. We are interested in stars of m$_V$ < 8 and effects on 
star counts of $\Delta$m$_V > \pm 0.1$. At this relatively bright limiting 
magnitude and amplitude range, the sample is essentially complete and still 
accounts for well less than one star per camera frame. Hence, the effect of 
variable stars biasing our sample is negligible compared with the star match error in m$_V$.
	
Throughput differences between the cameras, due largely to varying
sensitivities of the S-20 photocathodes, amount to $\Delta$m$_{V cathode}$ 
$\sim$ 0.2. This effect is significant but smaller than the star match 
magnitude error due to the predominance of the sky background.

We estimate our best-case scattered light level at about m$_V$ = 8.0. Our
resolution element angular size is 72 arcsec square, which corresponds to a
background level of m$_V$ = 10.2 per pixel for the best local
limiting magnitude of 19.5 per square arcsecond. Unfortunately, the instrument
was located near bright outdoor lighting, which increased the background by a
factor of several due to scattering by local dust and aerosols. Events
occurred at various angular distances from the Moon and at a variety of lunar
phases, significantly increasing the background for certain events. The data
show that both the limiting magnitude and the completeness of star matches
varied between gamma-ray bursts by $\Delta$m$_V$ $\sim$ 0.5. Similar results 
are expected for sky background-limited data.

In summary, $\Delta$m$_{V complete}$ $\cong$ m$_{V limit}$ $\cong$ 0.5 over 
all events is consistent with the sky background variations dominating our 
flux uncertainty values, albeit with instrumental noise and a slight star 
temperature uncertainty contribution as well.

\section{Connection Between Optical Limits and Gamma-ray Observations -
Discussion and Summary}

We want to relate the observed optical energy flux $F_\nu (\nu_o)$, (units of
ergs-cm$^{-2}$-s$^{-1}$-Hz$^{-1}$) at a fiducial frequency $\nu_o$ to the 
gamma-ray observations of the burst. For each burst, we have a 
series of optical upper limits from
images taken between t$_b$ and t$_e$. Clearly an optical upper limit is more
constraining for a bright GRB than for a weak one. The comparisons between the
two energy bands are model-dependent since the choice of optical and gamma-ray
quantities which are compared must be based explicitly or implicitly on an
assumed physical connection between these quantities. In general, we have
limits on the optical emission after the gamma-ray emission ends, while only
in a few cases do we have limits on the optical emission during the burst. Nor
can we say much about the bolometric optical emission because we do not know
what spectral shape we are constraining. The comparisons between gamma-ray and
optical quantities are not unique; in particular, they can be scaled by
constant factors such as fiducial energies and frequencies. Ultimately we
present quantities which can be used to constrain detailed physical models.
Our data support three different methods of comparing the optical and
gamma-ray observations.

Suppose first that optical emission is produced in the same region as the
gamma rays by related physical mechanisms. The optical flux then scales with
the gamma-ray photon flux at a fiducial energy E$_\gamma$, for example if 
the optical spectrum is an extrapolation of the gamma-ray spectrum 
(Ford \& Band 1996). For an event captured optically while the GRB 
was still in progress, the optical $F_\nu (\nu_o)$ should be compared 
to the gamma-ray energy flux F$_\gamma$(E$_\gamma$)
(ergs-cm$^{-2}$-s$^{-1}$-keV$^{-1}$) at E$_\gamma$(keV) averaged 
over the same time period as the image corresponding to F$_\nu$. This 
comparison suggests a simple ratio
\begin{equation}
R_1 = 2.42 \times 10^{17} {\hbox{Hz}\over \hbox{keV}} 
   {{F_\nu } \over {F_\gamma (E_\gamma) }}
\eqnum{2}
\end{equation}
or an effective energy spectral index
\begin{equation}
\alpha_{0 \gamma} = {\log \left ( 2.42\times 10^{17} 
{\hbox{Hz}\over \hbox{keV}} F_\nu / {F_\gamma (E_\gamma) } 
\right)  \over
\log \left (E_\gamma / {h\nu_0} \right)}
\eqnum{3}
\end{equation}
where E$_\gamma$ and  h$\nu_o$ must be expressed in the same units.

In the second case, optical emission may result from immediate reprocessing
of the gamma-ray flux. Then the optical flux at any moment during the burst
would be proportional to the gamma-ray energy flux (ergs-cm$^{-2}$-s$^{-1}$). 
If an optical image was accumulated from t$_b$ to t$_e$ = t$_b$ + $\Delta$t, 
F$_\nu$ should be compared to
\begin{equation}
\Phi_\gamma = {1\over {\Delta t}} \int^{t_e}_{t_b} dt \int dE\,\, F_\gamma(E,t)
\eqnum{4}
\end{equation}
where the integration proceeds over the model-dependent gamma-ray energy
range. A useful dimensionless quantity is
\begin{equation}
R_2 = {{F_\nu \nu_o} \over \Phi_\gamma} \quad.
\eqnum{5}
\end{equation}
In this expression, $\nu_o$ converts the optical energy flux per unit frequency
into a broadband energy flux. This quantity is approximately the inverse of
the Schaefer (1981) L$_\gamma$ / L$_{opt}$ integrated over the B-band.

Optical emission in a third case might be proportional to GRB fluence
following a time delay. The most important strength measure would be the total
GRB energy released, which is proportional to the burst fluence $\phi_\gamma$
observed at Earth. Thus the optical flux limits F$_\nu$ after the burst 
should be compared with $\phi_\gamma$. The ratio 
\begin{equation}
R_3 = {F_\nu \over \phi_\gamma}
\eqnum{6}
\end{equation}
is dimensionless (the units of F$_\nu$, ergs-cm$^{-2}$-s$^{-1}$-Hz$^{-1}$ 
include the dimensionless factor s$^{-1}$-Hz$^{-1}$). Note that optical 
R$_3$ could be multiplied by $\nu_o \Delta t$  and still
remain dimensionless. Mulitiplying F$_\nu$ by $\nu_o \Delta t$ would 
convert the energy flux per unit frequency into a fluence; formally, this 
latter form of R$_3$ compares
similar optical and gamma-ray quantities. We are instead interested in
describing the optical response at a given time to a burst of a specified
intensity so the characterization of the response a time after the burst
should be independent of the time over which an image is integrated. We thus
prefer to leave out this factor of $\nu_o \Delta t$ from the 
definition of R$_3$.

GROCSE often did not view the error box until the burst was over. The
comparison in such circumstances is best made between the optical upper limit
and the gamma-ray fluence, or R$_3$ (eq. 6). For those bursts where the optical
observation occurred while the burst was still in-progress, the gamma-ray
fluence up to the time of the optical observation should be used, although in
actuality both cases of a GROCSE observation occurring during a GRB caught
only the trailing edge of the burst after almost all the GRB energy emission.
 
Adequate data is made available to accommodate calculation of such
model-dependent temporal variations. Gamma-ray fluences can be derived from
several types of BATSE data products. We have the highest confidence in
fluences calculated by integrating the burst photon spectrum over energy and
time when the spectrum is fit to the SHERB datatype, which provides sufficient
spectral and temporal resolution. The SHERB data are a series of count spectra
accumulated by the BATSE Spectroscopy Detectors (SDs); except for the longest
bursts (none of which were observed by GROCSE), there are SHERB spectra after
the end of each burst which help in extrapolating the background spectrum
during the burst.

When there are no SHERB spectra for one reason or another (e.g., the SHERB
spectra are not returned in the telemetry for weak bursts, or data are lost in
telemetry gaps or gaps in collection), we used the fluences from the 3B
catalog (Meegan et al. 1996), when available. These fluences are provided over
the 20-2000 keV range. Finally, we used fluences derived from fits to the STTE
spectra which are also from the SDs. The STTE data are the arrival times and
energies of 64,000 counts around the time of the burst trigger from various
detector modules; these counts can be accumulated over a variety of time
scales. The STTE data do not necessarily span the entire burst duration nor do
they provide sufficient background data to assist in interpolating its
background spectra at the time of the burst. Thus we use the STTE data as a
last resort. 
	
The results of this comparison are presented in columns 8 through 10 of Table
1. Column 8 describes the gamma-ray fluence $\phi_\gamma$ in units of
ergs-cm$^{-2}$
over the energy range 20-2000 keV. The ratio of optical flux to gamma-ray
fluence R$_3$, the ratio of the optical flux upper limit to the gamma-ray 
fluence is shown in column 9. Note that this ratio is an upper limit. Column 
10 provides notes as applicable to each burst. Table 2 shows values of $\Phi$, 
R$_1$, and R$_2$ for selected GRBs, as defined in equations (4), (2), and (5),
respectively for simultaneous optical gamma-ray emission. Note that the upper
limits are given for the latter two quantities.

If the optical limit was set during the gamma-ray burst, upper limits for
in-band gamma-ray flux can be computed based on extrapolating eq. (2) to
visible wavelengths. Burst spectra can be fitted with the "GRB" spectral
function (Band et al. 1993). Here, the gamma-ray energy spectra are modeled in
terms of two components: a high-energy component and a low-energy component of
functional form
\begin{equation}
N(E) = {{F_\gamma}\over {E_\gamma}} = AE^\alpha e^{-E/E_o} \, ,  \quad 
E \le (\alpha-\beta)E_o
\eqnum{7a}
\end{equation}
for low energies and 
\begin{equation}
N(E) = BE^\beta \, ,  \quad E > (\alpha-\beta)E_o 
\eqnum{7b}
\end{equation}
for high energies, where the parameters A, $\alpha$, B, and $\beta$ vary 
amongst gamma-ray bursts, and $\alpha > \beta$. The energy breakpoint is 
chosen so that 
N(E) and its derivative are continuous at E$_o$, typically 100 keV to $>$1 MeV.

Fortuitously, several bursts appeared to be in progress at the moment of
GROCSE imaging, among them GRB 951220 and 951124. There are no SHERB data
for GRB 951220 and the STTE data are inadequate to determine the gamma-ray
flux during the optical observation. Thus, an energy flux from GRB 951220 may
be computed from observations although insufficient data exist to reconstruct
a spectrum. The GRB 951124 optical observation occurred on the falling edge of
the first of two gamma-ray flux peaks. Based on the evolution of the spectrum
as gleaned from SHERB data, 90\% of the energy flux resides in the first peak.
Flux results up to the optical observation and for the entire event are
presented in Table 1. The GROCSE observation began 30 s after the BACODINE
trigger. The SHERB data were accumulated at the following post-trigger times: 
26.75, 28.8, 31.30, \& 34.30 s. Due to the small number of counts and
variable count rate, as expected from a highly dynamic spectral evolution,
determination of the instantaneous spectrum at t = 30.0 s is somewhat
imprecise. The best fit is:
\begin{equation}
N(E) = 8.413 \times 10^{-3} \left({{E}\over{100\hbox{ keV}}}\right)^{-0.3464}
   e^{-E/85.9} \quad , \quad E\le 177.75\hbox{ keV},
\eqnum{8a}
\end{equation}
\begin{equation}
N(E) = 3.493 \times 10^{-3} \left({{E}\over{100\hbox{ keV}}}\right)^{-2.416}
   \quad , \quad \quad \quad \quad \quad E > 177.75\hbox{ keV}.
\eqnum{8b}
\end{equation}
In conclusion GROCSE did not observe optical counterparts to GRBs to levels
described in Table 2. Proper physical mechanisms for GRBs must not yield
optical components larger than these new upper limits.

GROCSE was decommissioned in June, 1996 following two years of successful
operation, in order to concentrate on the next-generation instrument LOTIS, a
fast-slewing wide-field lens array featuring four Loral 2048 $\times$ 2048 CCDs
capable of searching to below m$_V$ = 14. LOTIS is now fully-operational and
results from this new experiment will be presented in a forthcoming paper.

This work was supported by the U.S. Department of  Energy, under contract
W-7405-ENG-48 to the Lawrence Livermore National Laboratory and NASA contract
S-57771-F. Gamma-ray burst research at UCSD  (D. Band) is supported by NASA 
contract NAS8-36081. K. Hurley acknowledges JPL Contract 958056 for Ulysses 
operations and NASA Grant NAG5-1560 for IPN work. Brian Lee is grateful for 
the support of his work by grant NGT-52805 under the NASA Graduate Student 
Researchers Program. We also acknowledge C. Akerlof from Univ. of Michigan 
for his participation of this experiment.

\clearpage



\begin{deluxetable}{crrrrrrrrrrr}
\tightenlines
\scriptsize
\tablecaption{Summary of GROCSE Event Characteristics. \label{tbl-1}}
\tablewidth{0pt}
\tablehead {
\colhead{Name}         & \colhead{BATSE}             & 
\colhead{m$_V$}        & \colhead{F$_\nu(5500 \AA)^a$} & 
\colhead{t$_d^b$}      & \colhead{T$_{90}^c$}        & 
\colhead{Cov$^d$}      & \colhead{F$_\gamma^e$}      & 
\colhead{R$_3$}        & \colhead{notes} \\
\colhead{}             & \colhead{Trigger}           &
\colhead{V}            & \colhead{} 		   &
\colhead{s}            & \colhead{s}                 &
\colhead{\%}           & \colhead{} 		   &
\colhead{}             & \colhead{}       
}
 
\startdata
940129	&2793	&7.3	&4.58$\times 10^{-23}$	&35	&7	&75
	&1.53$\times 10^{-5}$	&2.94$\times 10^{-18}$	&f,k	\\
940623	&3040	&7.3	&4.58$\times 10^{-23}$	&17.4	&26	&80
	&2.88$\times 10^{-6}$	&1.56$\times 10^{-17}$	&g	\\
	&	&7.3	&4.58$\times 10^{-23}$	&20.3	&	&
	&3.10$\times 10^{-6}$	&1.45$\times 10^{-17}$	&g	\\
	&	&7.3	&4.58$\times 10^{-23}$	&24.1	&	&
	&3.28$\times 10^{-6}$	&1.37$\times 10^{-17}$	&g	\\
	&	&7.3	&4.58$\times 10^{-23}$	&32	&	&
	&3.38$\times 10^{-6}$	&1.33$\times 10^{-17}$	&g	\\
940828	&3141	&8.5	&1.52$\times 10^{-23}$	&21	&2.3	&50
	&2.20$\times 10^{-6}$	&6.77$\times 10^{-18}$	&f,i	\\
940907	&3159	&7.0	&6.04$\times 10^{-23}$	&22	&18.2	&25
	&1.42$\times 10^{-6}$	&4.18$\times 10^{-17}$	&g	\\
950531	&3611	&7.1	&5.51$\times 10^{-23}$	&23	&3	&80
	&2.44$\times 10^{-7}$	&2.22$\times 10^{-16}$	&j	\\
950907	&3779	&7.5	&3.81$\times 10^{-23}$	&35	&7	&80
	&3.89$\times 10^{-7}$	&9.63$\times 10^{-17}$	&j	\\
950918	&3805	&7.7	&3.17$\times 10^{-23}$	&20	&40	&90
	&8.21$\times 10^{-7}$	&3.80$\times 10^{-17}$	&g	\\
950922	&3814	&7.0	&6.04$\times 10^{-23}$	&46	&5	&90
	&1.04$\times 10^{-6}$	&5.71$\times 10^{-17}$	&j	\\
951117	&3909	&7.0	&2.40$\times 10^{-23}$	&18	&25	&90
	&2.86$\times 10^{-6}$	&2.08$\times 10^{-17}$	&g,k	\\
	&	&7.0	&2.40$\times 10^{-23}$	&25	&	&
	&3.15$\times 10^{-6}$	&1.89$\times 10^{-17}$	&g	\\
951124	&3918	&7.4	&4.18$\times 10^{-23}$	&23.1	&150	&85
	&1.33$\times 10^{-5}$	&3.09$\times 10^{-18}$	&g,k	\\
	&	&7.4	&4.18$\times 10^{-23}$	&29.5	&	&
	&1.56$\times 10^{-5}$	&2.63$\times 10^{-18}$	&g	\\
	&	&7.4	&4.18$\times 10^{-23}$	&64.1	&	&
	&1.68$\times 10^{-5}$	&2.45$\times 10^{-18}$	&g	\\
	&	&7.4	&4.18$\times 10^{-23}$	&69.0	&	&
	&1.68$\times 10^{-5}$	&2.45$\times 10^{-18}$	&g	\\
	&	&7.4	&4.18$\times 10^{-23}$	&91.5	&	&
	&1.68$\times 10^{-5}$	&2.45$\times 10^{-18}$	&g	\\
	&	&7.4	&4.18$\times 10^{-23}$	&96.4	&	&
	&1.68$\times 10^{-5}$	&2.45$\times 10^{-18}$	&g	\\
	&	&7.4	&4.18$\times 10^{-23}$	&119.4	&	&
	&1.86$\times 10^{-5}$	&2.21$\times 10^{-18}$	&g	\\
	&	&7.4	&4.18$\times 10^{-23}$	&124.3	&	&
	&1.86$\times 10^{-5}$	&2.17$\times 10^{-18}$	&g	\\
	&	&7.4	&4.18$\times 10^{-23}$	&146.8	&	&
	&1.89$\times 10^{-5}$	&2.17$\times 10^{-18}$	&g	\\
	&	&7.4	&4.18$\times 10^{-23}$	&151.7	&	&
	&1.89$\times 10^{-5}$	&2.22$\times 10^{-18}$	&g	\\
951208	&3936	&7.0	&6.04$\times 10^{-23}$	&20	&3.5	&85
	&2.67$\times 10^{-6}$	&2.22$\times 10^{-17}$	&h,k	\\
951220	&4048	&7.9	&2.64$\times 10^{-23}$	&14.8	&17	&95
	&1.24$\times 10^{-5}$	&2.09$\times 10^{-18}$	&g,k	\\
960403	&5407	&7.0	&6.04$\times 10^{-23}$	&33	&70	&20
	&3.12$\times 10^{-6}$	&1.90$\times 10^{-17}$	&h	\\
 
\enddata

 
\tablenotetext{a}{Optical flux (ergs-cm$^{-2}$-s$^{-1}$-Hz$^{-1}$)}
\tablenotetext{b}{Time in seconds between burst trigger and
GROCSE observation}
\tablenotetext{c}{GRB duration in seconds}
\tablenotetext{d}{GROCSE coverage of 3-$\sigma$ BATSE error box}
\tablenotetext{e}{20-2000 keV gamma-ray fluence (ergs-cm$^{-2}$)}
\tablenotetext{f}{Fluence (erg-cm$^{-2}$) from 3rd BATSE catalog (Meegan, 1996)}
\tablenotetext{g}{Fluence calculated from spectral fit to HERB data}
\tablenotetext{h}{Fluence calculated from spectral fit to SHERB data}
\tablenotetext{i}{Data missing}
\tablenotetext{j}{Fluence calculated from spectral fit to STTE data, 
background subtraction uncertain}
\tablenotetext{k}{IPN annulus available}
 
\end{deluxetable}


\clearpage



\begin{deluxetable}{crrrrrrrrrrr}
\tightenlines
\scriptsize
\tablecaption{Calculation of Comparison Quantities for Selected Observations. \label{tbl-2}}
\tablewidth{0pt}
\tablehead {
\colhead{Name}			& \colhead{BATSE}			& 
\colhead{t$_d^a$}		& \colhead{F$_\gamma(100 keV)^b$}	& 
\colhead{$\Phi^c_\gamma$}	& \colhead{R$_1$}			& 
\colhead{R$_2$}\\
\colhead{}			& \colhead{Trigger}		&
\colhead{s}			& \colhead{}			&
\colhead{}			& \colhead{upper limit}		&
\colhead{upper limit}
}
 
\startdata
940623	&3040	&17.4	&2.42 $\times 10^{-10}$	&9.28 $\times 10^{-8}$	
	&4.50 $\times 10^{4}$	&2.91 $\times 10^{-1}$ \\
	&	&20.3	&1.77 $\times 10^{-10}$	&6.26 $\times 10^{-8}$	
	&6.15 $\times 10^{4}$	&4.32 $\times 10^{-1}$\\
951117	&3909	&18	&1.72 $\times 10^{-10}$	&7.55 $\times 10^{-8}$	
	&8.34 $\times 10^{4}$	&4.72 $\times 10^{-1}$ \\
951124	&3918	&23.1	&8.89 $\times 10^{-10}$	&5.94 $\times 10^{-7}$	
	&1.12 $\times 10^{4}$	&4.15 $\times 10^{-2}$ \\
	&	&29.5	&4.46 $\times 10^{-10}$	&1.64 $\times 10^{-7}$	
	&2.23 $\times 10^{4}$	&1.50 $\times 10^{-1}$ \\
	&	&119.4	&1.45 $\times 10^{-10}$	&1.03 $\times 10^{-7}$	
	&6.85 $\times 10^{4}$	&2.39 $\times 10^{-1}$ \\
	&	&124.3	&9.66 $\times 10^{-12}$	&4.65 $\times 10^{-9}$	
	&1.03 $\times 10^{6}$	&5.30 \\
951220	&4048	&14.8	&4.12 $\times 10^{-10}$	&2.66 $\times 10^{-7}$	
	&1.52 $\times 10^{4}$	&5.84 $\times 10^{-2}$
 
\enddata

 
\tablenotetext{a}{Time in seconds between burst trigger and
GROCSE observation}
\tablenotetext{b}{Gamma-ray flux at 100 keV (ergs-cm$^{-2}$-s$^{-1}$-keV$^{-1}$)}
\tablenotetext{c}{Gamma-ray fluence (erg-cm$^{-2}$-s$^{-1}$)}
 
\end{deluxetable}



\clearpage

\clearpage

\begin{figure}
\plotone{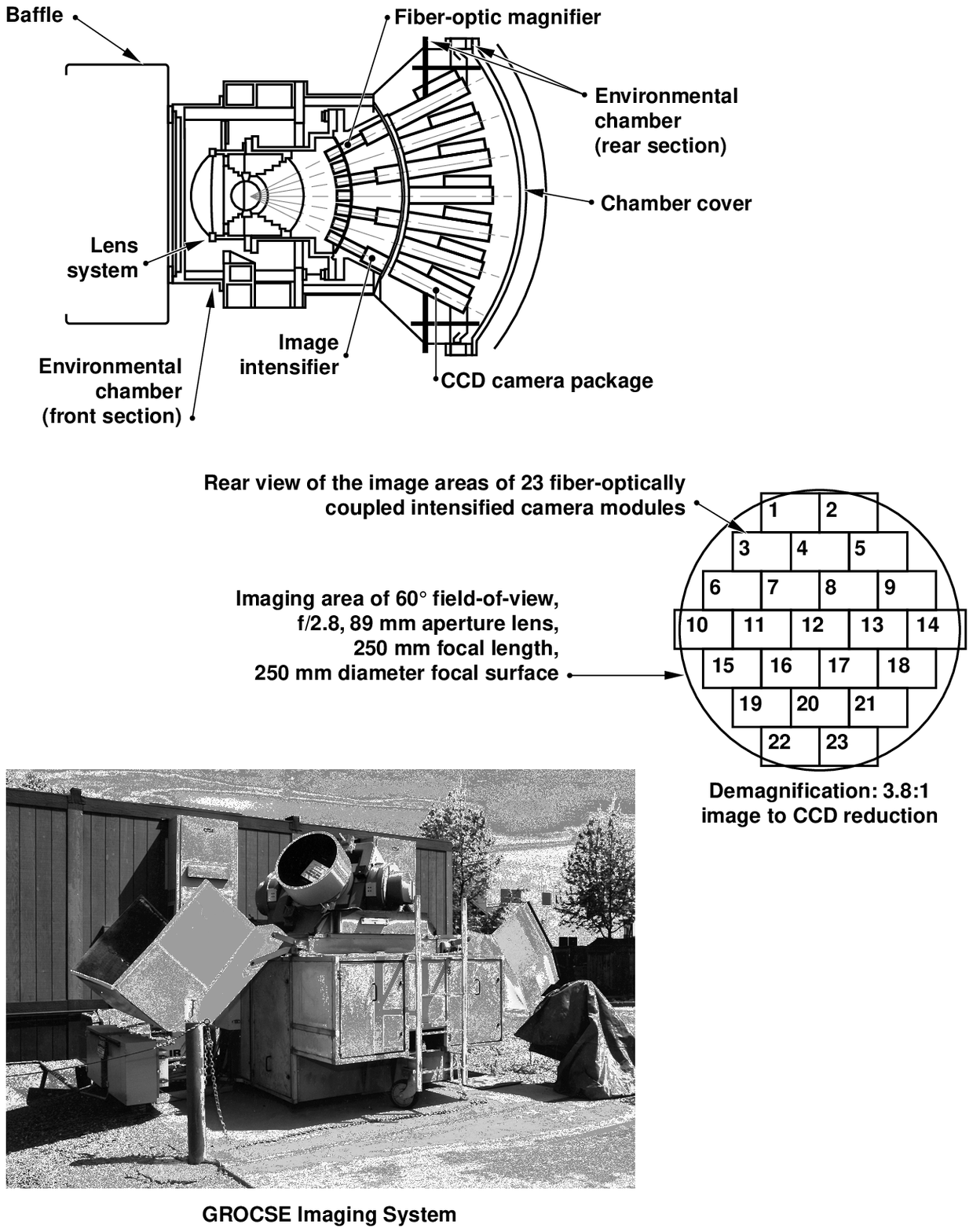}
\caption{The GROCSE system showing the fish-eye lens with fiber optic 
couplers mounted on the curved focal surface. Each fiber optic bundle 
for each camera is cemented to an intensifier and then imaged onto a CCD. 
The lay out of the 23 cameras viewing 60$^\circ$ field is depicted in the 
second panel. The entire
instrument is mounted on a heavy-duty rapidly slewing alt-zimuth mount. 
The electronic 
interface and computer control hardware for this fully-automated system are 
not visible. \label{fig1}}
\end{figure}

\clearpage
\begin{figure}
\plotone{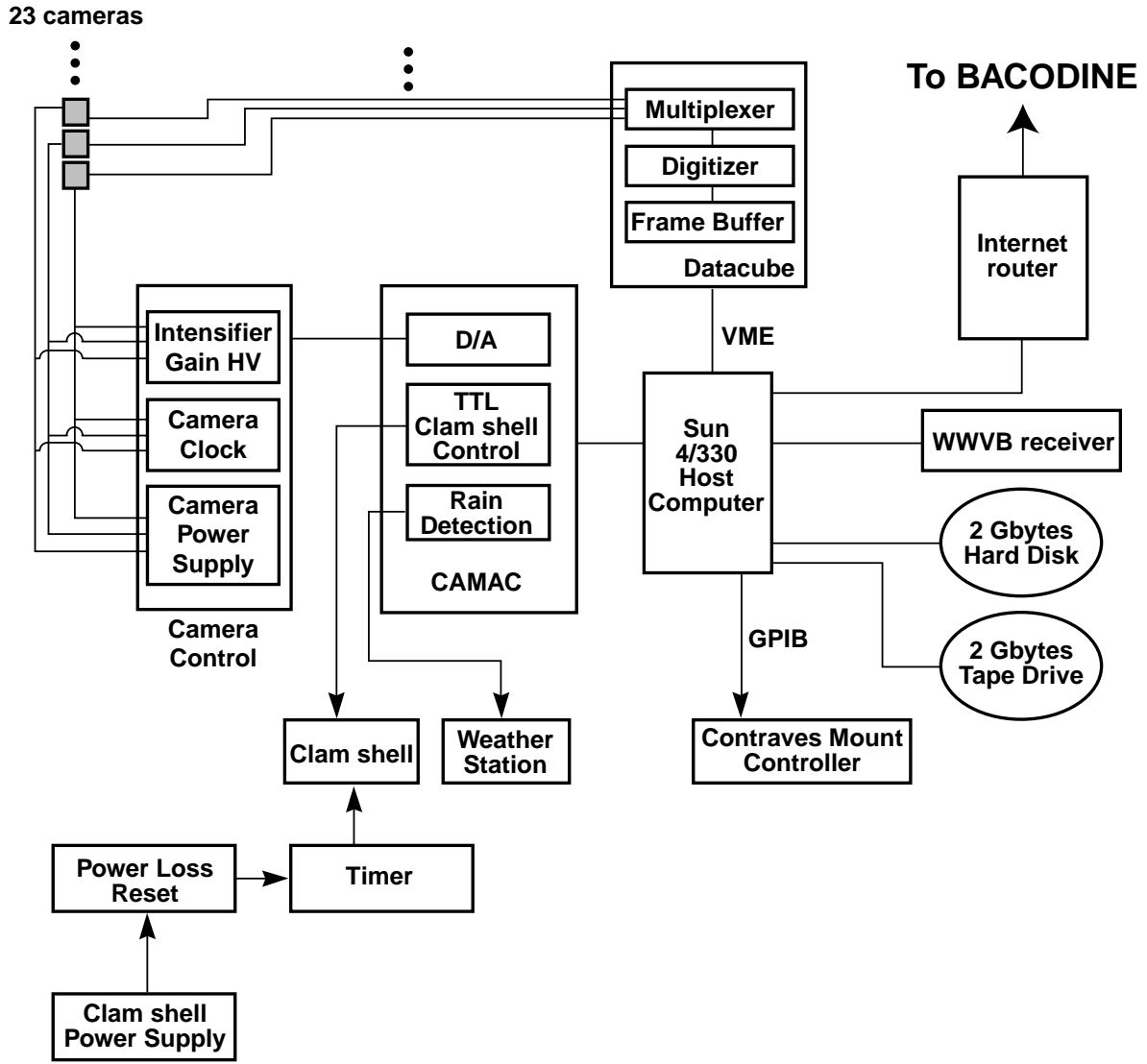}
\caption{Schematic drawing of the GROCSE data acquisition system. \label{fig2}}
\end{figure}

\clearpage
\begin{figure}
\plotone{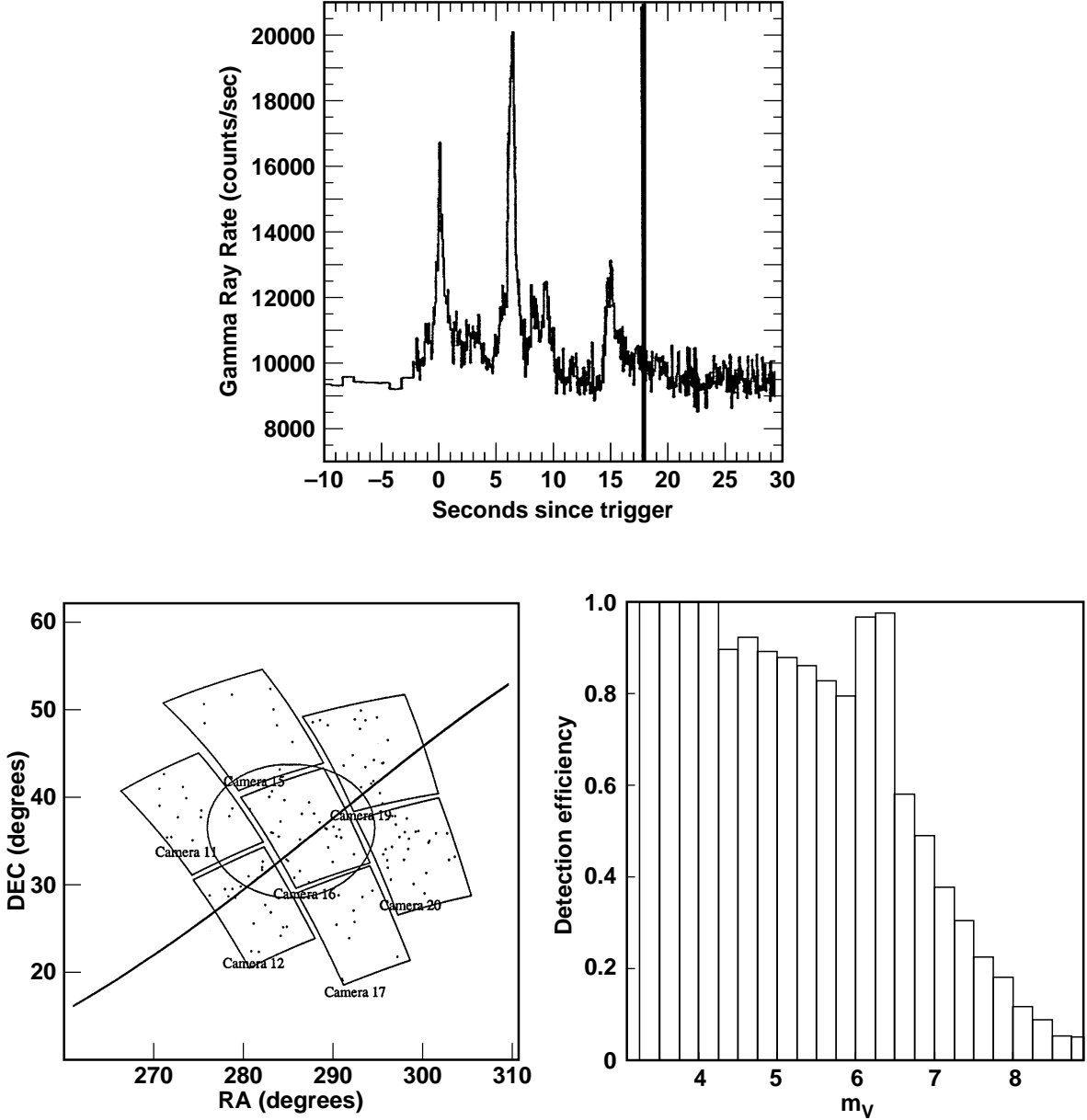}
\caption{GRB 951117 (BATSE Trigger 3909) event. The first panel is the BATSE
light curve. The BATSE trigger occurred at time zero. Shown on the ordinate is
the count rate in counts/s. The vertical line represents the instant when
GROCSE imaging began. The second panel is the GROCSE coverage of this event.
The 3-$\sigma$ BATSE error circle and the IPN annulus are superimposed on the
GROCSE rectangular field. All GROCSE image exposure times were 0.50 s. The
third panel shows the star count versus magnitude. Note that the completeness
magnitude is defined as the level for which all stars are observed while the
limiting magnitude is the brightness where only half the stars are observed.
\label{fig3}}
\end{figure}

\clearpage
\begin{figure}
\plotone{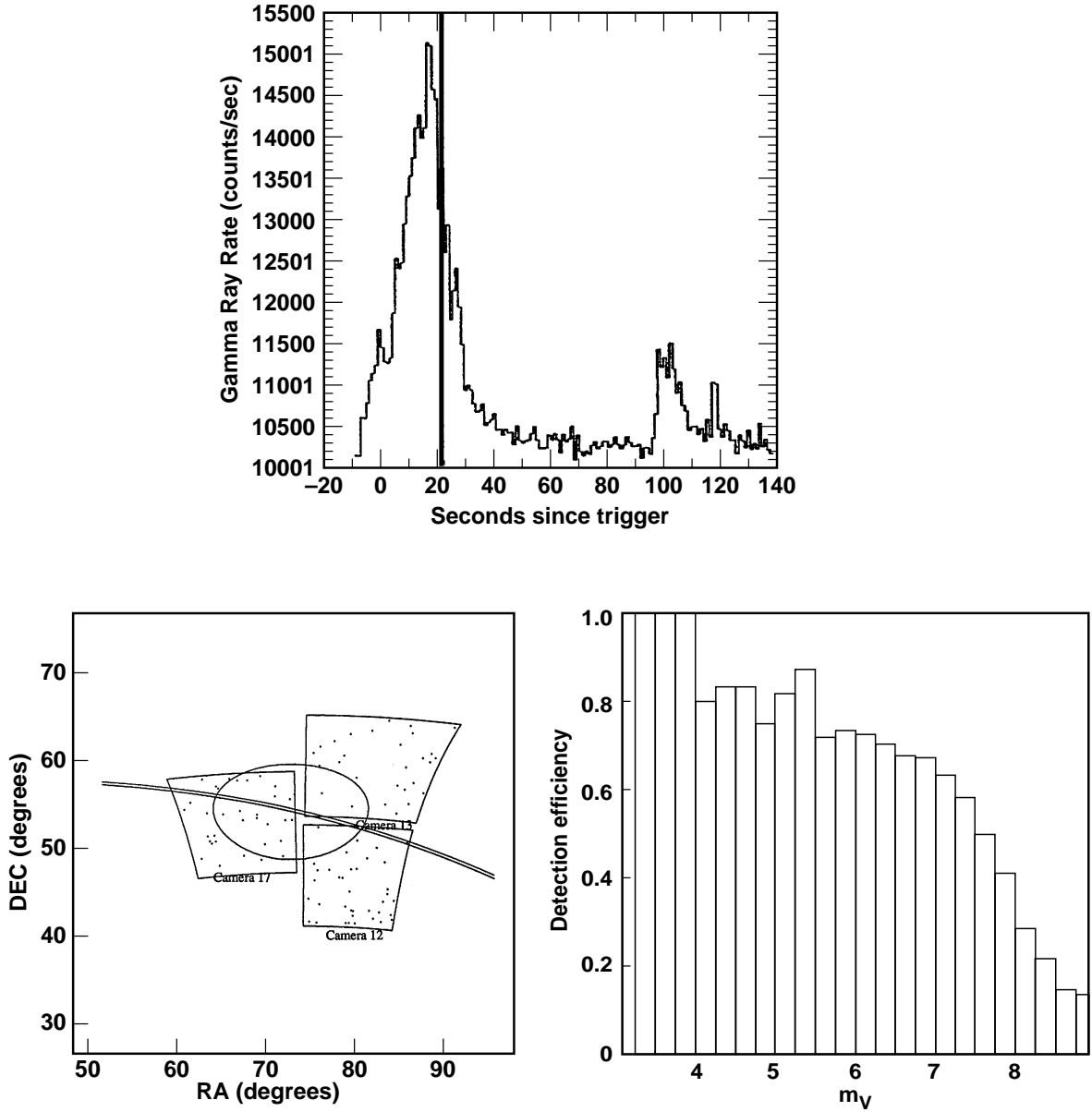}
\caption{GRB 951124 (BATSE Trigger 3918) event. \label{fig4}}
\end{figure}

\clearpage
\begin{figure}
\plotone{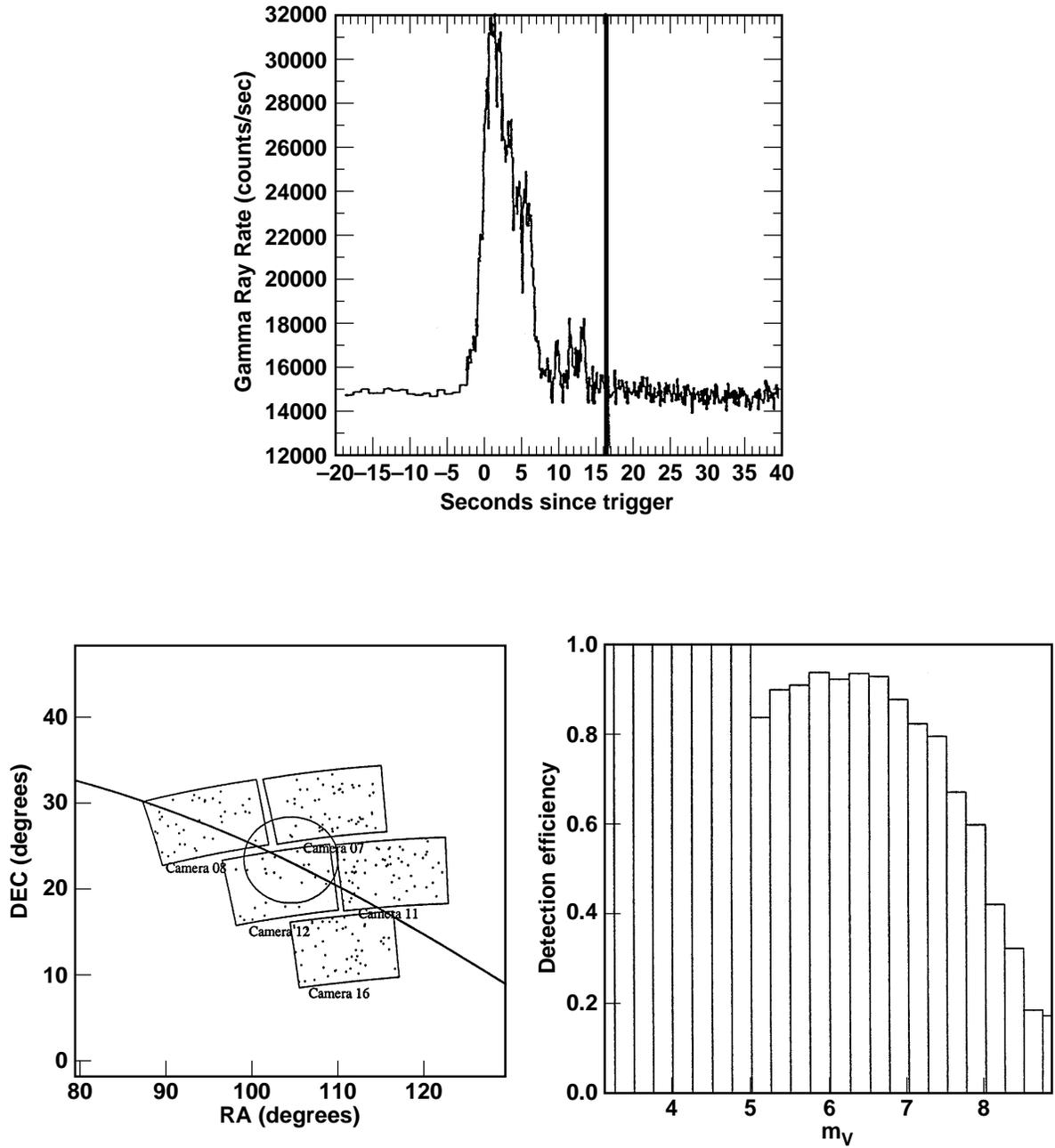}
\caption{GRB 951220 (BATSE Trigger 4048) event. \label{fig5}}
\end{figure}



%

\end{document}